\documentclass[superscriptaddress,aps,prb,twocolumn]{revtex4-1}
\usepackage{graphicx}
\usepackage{bm}
\usepackage{amsmath,amssymb,amsfonts}
\usepackage{epsfig}
\usepackage{epstopdf}
\usepackage{dcolumn}
\usepackage{grffile}
\usepackage{verbatim}
\usepackage{mathrsfs}
\usepackage{xr}
\usepackage[colorlinks=true,linkcolor=blue,citecolor=blue, urlcolor=blue]{hyperref}

\begin{document}
\title{t-J model on the effective brick-wall lattice for the recently discovered high-temperature superconductor Ba$_2$CuO$_{3+\delta}$}

\author{Zhan Wang}
\affiliation{Kavli Institute for Theoretical Sciences, University of Chinese Academy of Sciences, Beijing 100190, China}

\author{Sen Zhou}
\email{zhousen@itp.ac.cn}
\affiliation{Institute of Theoretical Physics, Chinese Academy of Sciences, Beijing, 100190, China}
\affiliation{Center for Excellence for Topological Quantum Computation, Chinese Academy of Sciences, Beijing, 100190, China}

\author{Weiqiang Chen}
\email{chenwq@sustech.edu.cn}
\affiliation{Institute for Quantum Science and Engineering and Department of Physics, Southern University of Science and Technology, Shenzhen, 518055, China}

\author{Fu-Chun Zhang}
\email{fuchun@ucas.ac.cn}
\affiliation{Kavli Institute for Theoretical Sciences, University of Chinese Academy of Sciences, Beijing 100190, China}
\affiliation{Center for Excellence for Topological Quantum Computation, Chinese Academy of Sciences, Beijing, 100190, China}

\date{\today}

\begin{abstract}

Layered copper oxides have highest superconducting transition temperatures at ambient pressure. Its mechanism remains a grand challenge in condensed matter physics.  The essential physics lying in 2-dimensional copper-oxygen layers is well described by a single band Hubbard model or its strong coupling limit t-J model in 2-dimensional square lattice.  Recently discovered high temperature superconductor Ba$_2$CuO$_{3+\delta}$ with $\delta \sim 0.2$ has different crystal structure with large portion of in-plane oxygen vacancies. We observe that an oxygen vacancy breaks the bond of its two neighboring copper atoms, and propose ordered vacancies in Ba$_2$CuO$_{3+\delta}$ lead to extended t-J model on an effective brick-wall lattice. For the nearest neighbor hopping, the brick-wall model can be mapped onto t-J model on honeycomb lattice. Our theory explains the superconductivity of Ba$_2$CuO$_{3+\delta}$ at high charge carrier density, and predict a time reversal symmetry broken pairing state.

\end{abstract}
\maketitle

\emph{Introduction.} Understanding of high Tc copper oxides\cite{Keimer2015, Anderson2004, Rice2011, Bednorz1986, RevModPhys.78.17} remains a grand challenge in condensed matter physics despite intensive studies in the past over thirty years.  It is generally accepted that the basic physics of high $T_c$ cuprates is on $\text{CuO}_2$ planes, and the half-filled $d_{x^2-y^2}$ holes of formal $\text{Cu}^{2+}$ form Mott insulator or charge transfer insulator. Chemical doping to the parent compounds introduces additional holes and leads to the $d$-wave superconductivity with the highest transition temperatures at ambient pressure. Recently, the superconductivity is discovered in $\text{Ba}_2\text{CuO}_{3+\delta}$ material with $\delta \approx 0.2$, which belongs to a different family of high $T_c$ cuprate\cite{Li12156,Scalapino2019}. The Meissner effect measurement shows that the compound has transition temperature $T_c = 73K$, which is much higher than that of the similar structure high $T_c$ cuprate $\text{Ba}_x\text{La}_{1-x}\text{CuO}_4$. It is then interesting and important to examine the electronic structure and superconductivity of $\text{Ba}_2\text{CuO}_{3+\delta}$, which may also shed new light on our understanding of high $T_c$ cuprates in general.

$\text{Ba}_2\text{CuO}_{3+\delta}$ has compressed octahedron with the apical oxygen atoms closer to the central Cu-atoms, the Cu-O bond length for the apical O is about 1.9 \AA, shorter than the Cu-O bond length on the Cu-O plane. This is different from all other known cuprates, where the Cu-O bond along the apical O is longer. There are a large number of O-vacancies in $\text{Ba}_2\text{CuO}_{3+\delta}$, which are located on the $\text{CuO}_2$ plane as neutron data showed \footnote{Private communication with Q. Z. Huang and C. Q. Jin.}. These vacancies are expected to play important role in the electron structure and possibly superconductivity. First principles calculations have been reported by Liu et al\cite{PhysRevMaterials.3.044802}, who considered various possible crystal structures. For $\delta=0.25$, they have found crystal structures corresponding to lower total energies. There have been a number of proposed theories for its electronic structure and superconductivity. In the usual cuprates, the formal valence of Cu is less than $+2.2$, corresponding to the hole concentration on the $\text{CuO}_2$ plane less than $20\%$.  In $\text{Ba}_2\text{CuO}_{3+\delta}$, the average valence of Cu is $+2(1+\delta)$, corresponding to average hole concentration of $2\delta \gg 0.2$ for the value of given $\delta\sim0.2$.  This has led to the proposal of 2-band Hubbard model to describe the compound\cite{PhysRevB.99.224515,2019arXiv190912620L}. Large number of O vacancies has led to the proposal of weakly linked 2-chain ladder for the superconductivity\cite{2019arXiv190908304L}, whose related physics has been previously examined extensively in study of Hubbard or t-J ladders\cite{Dagotto1996}.

\begin{figure}
\centering
\includegraphics[width=\columnwidth]{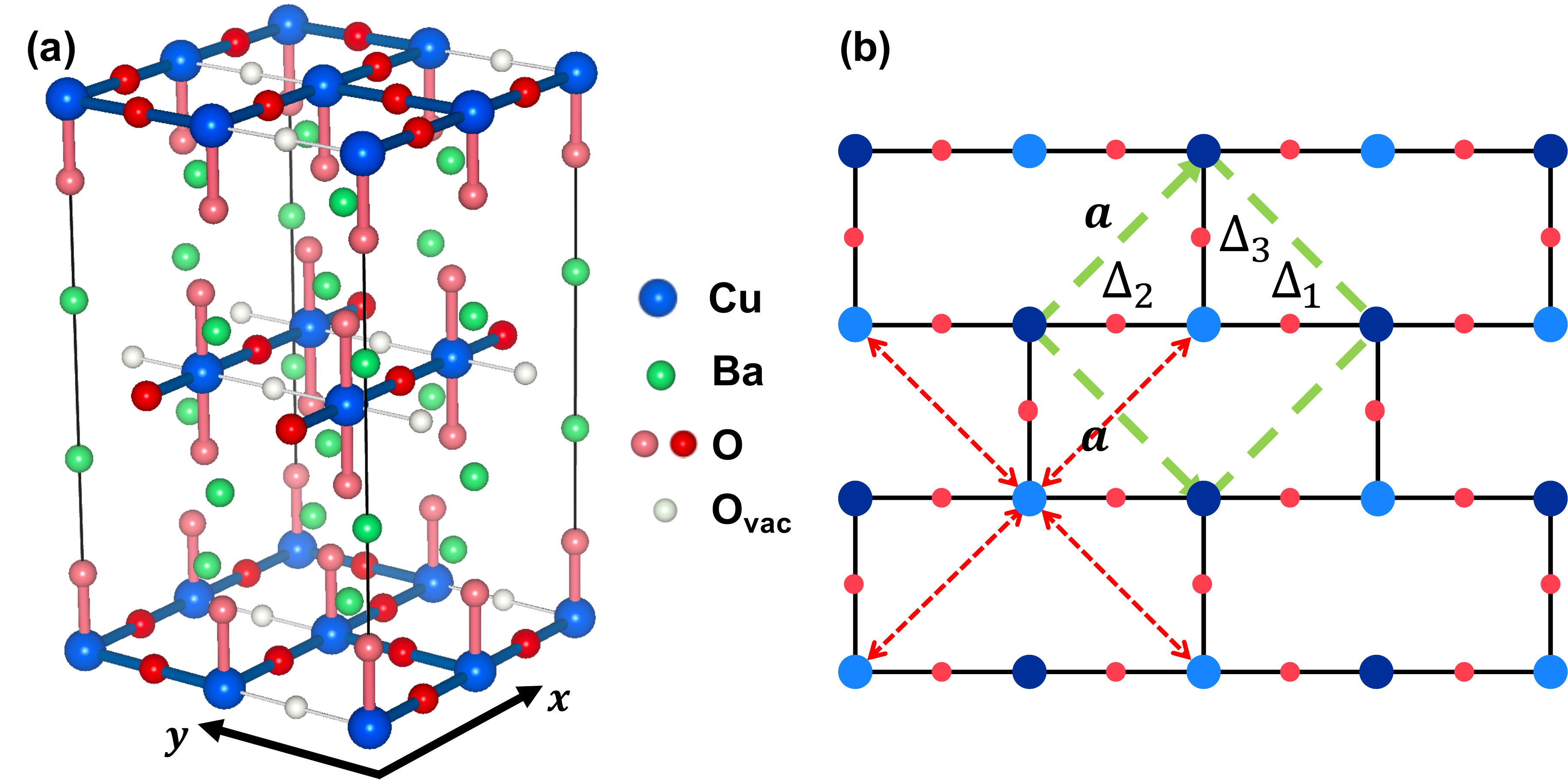}
\caption{(a) Proposed crystal structure for $\text{Ba}_2\text{CuO}_{3.25}$ with alternating Cu-O chain layer (middle layer) and Cu-$\text{O}_{1.5}$ plane (top and bottom layers). The apical O-atoms are all occupied. The Cu-$\text{O}_{1.5}$ plane forms a brick-wall structure due to the missing O-atoms. (b)The brick-wall lattice formed by Cu (light blue for A sublattice and dark blue for B sublattice) and O (red) atoms. $\Delta_\alpha,\alpha=1,2,3$ are the superconducting order parameters on the corresponding bonds. The unit cell vectors are depicted as green dashed lines with length $a$, while the next nearest neighbors are indicates by the red dashed arrows.}
\label{model}
\end{figure}

In this paper, we take the viewpoint that the observed superconductivity in $\text{Ba}_2\text{CuO}_{3+\delta}$ is resulted from an ordered crystal structure and the O-vacancy effectively transforms the original square lattice of $\text{CuO}_2$ to a different lattice. We propose a crystal structure for $\text{Ba}_2\text{CuO}_{3+\delta}$, where layers of 1-dimensional CuO chains and layers of 2-dimensional $\text{CuO}_{1.5}$ plane are alternating as illustrated in Fig.~\ref{model}(a). We expect the CuO chains to serve as charge reservoir, in analogy to the CuO chains in YBCO, and the CuO$_{1.5}$ planes to contain the essential physics for the superconductivity in Ba$_2$CuO$_{3+\delta}$. We argue that the effective Hamiltonian of a CuO$_{1.5}$ plane is described by a single band $t$-$t'$-$J$ model on an underlying brick-wall lattice, which is shown in fig. \ref{model}(b), with $t$ and $t'$ the nearest neighbor (n.n.) and next n.n. hopping integrals and $J$ the n.n. spin-spin coupling of the spin-1/2 Cu-holes.

We use renormalized mean field theory\cite{Zhang_1988} to study the $t$-$t'$-$J$ model on the brick-wall lattice. We find that the superconductivity extends to hole density of as high as $40\%$, and the maximum of the pairing order parameter appears at a larger hole concentration than the other cuprates due to the shift of the van Hove singularity for the density of states in the brick-wall lattice.  The pairing symmetry depends on the value of $t'/t$ and the hole concentration, and may break time reversal symmetry, which can be tested in muon spin rotation ($\mu$SR) experiment. The effects of bond disorder is also studied, and the superconductivity is expected to survive a weak bond disorder that deviates the structure from perfect brick-wall lattice.

Our theory appears to be consistent with the available experiments and explains the superconductivity of $\text{Ba}_2\text{CuO}_{3+\delta}$ with relatively large hole concentration. We attribute the change of the electronic structure in the planar layer of the compound to the O-vacancy, which in turn changes the effective lattice. This may provide another route to study high $T_c$ in future.

\emph{Model and effective lattice.} As mentioned above, we consider a crystal structure shown in Fig.~\ref{model} (a), with alternative layers of $\text{CuO}$ chain and $\text{CuO}_{1.5}$ plane (x-y plane). The average $\delta$ is $0.25$ in such a structure. The $\text{O}$-vacancies (missing O-atoms) in $\text{CuO}_{1.5}$ layer form a square lattice and there are three O-atoms around each $\text{Cu}$-atom on the plane. The energy of the crystal structure has not been calculated, but should be similar to the one of the lowest energy crystal structure calculated for $\text{Ba}_2\text{CuO}_{3.25}$ by Liu. et al\cite{PhysRevMaterials.3.044802}, with the difference that in the latter case the $\text{O}$-vacancies form an alternative chains along the x-direction. We will focus on the electronic structure of $\text{CuO}_{1.5}$ plane and consider CuO chains to serve as charge reservoir. While the average of the formal valence of Cu in $\text{Ba}_2\text{CuO}_{3.25}$ is $\text{Cu}^{2.5+}$, the formal valence of Cu-atom on the $\text{CuO}_{1.5}$ layer can be significantly smaller because of the compensation from the CuO chains. 

For each Cu-atom on the $\text{CuO}_{1.5}$ plane, there are five nearest neighbor $\text{O}$ atoms including two apical O-atoms, two on the x-axis and one on the y-axis.  The possible anti-bonding Cu-$3d$ orbitals here are $3d_{3z^2-r^2}$ or $3d_{z^2-x^2}$, whose relative energies depend on the difference or ratio of the bond lengths between the short apical Cu-O bond and the in-plane Cu-O bond. Here we assume orbital $3d_{3z^2-r^2}$ to have higher energy, which is consistent with a density functional calculation for proper parameters\footnote{Private communication with Congcong Le}. Then the lowest-energy of the atomic $3d$ hole state is $d_{3z^2-r^2}$, which replaces $3d_{x^2-y^2}$ in other cuprates as the relevant orbital. Considering Cu-$3d^{10}$ and O-$2p^6$ as the vacuum configuration, the formal $\text{Cu}^{2+}$ thus has one hole primarily on Cu-$3d_{3z^2-r^2}$. Because of the large repulsive interaction $U$ for two holes on the same Cu-site, the ground state at the half filled, namely one hole per Cu-atom in average, will be a Mott insulator, similar to that of $\text{La}_2\text{CuO}_4$.

We next consider additional holes to the half filled case.  As soft-X ray absorption experiment\cite{Li12156} shows, the additional hole largely goes to the O-2$p$ orbitals, which implies that the formal $\text{Cu}^{3+}$ has one hole on Cu-$3d_{3z^2-r^2}$ and the other on O-$2p$. We expect them to form a spin singlet, similar to the Zhang-Rice spin singlet\cite{PhysRevB.37.3759} formed in other cuprates, and moves through the lattice as a charge carrier.

Since the hopping of the hole between the two neighboring Cu-atoms is essentially mediated by oxygen $2p$ orbitals in between, it is strongly suppressed if the oxygen between the two Cu-atoms is missing. Therefore, the nearest neighbor Cu-Cu bond may be effectively removed from the lattice if the O-atom between them is missing.  This leads to an effective brick-wall lattice as shown in fig.\ref{model}(b).

As we discussed above, the physics in Ba$_2$CuO$_{3+\delta}$ may be similar to that in other cuprates and we may consider the material as a doped Mott insulator within a single orbital model, although the relevant Cu-$3d$ orbital and the underlying lattice due to the missing O ions will be different from the other cuprates. The low energy physics of $\text{Ba}_2\text{CuO}_{3.25}$ is thus described by a 2-dimensional t-J model on a brick-wall lattice,
\begin{equation}
\label{eq:1}
  \mathcal{H} = - \sum_{ij} \left(t_{ij} P_G c_{i \sigma}^{\dag} c_{j \sigma} P_G +\text{h.c.}\right)+ J\sum_{\left\langle ij \right\rangle} \mathbf{S}_i \cdot \mathbf{S}_j,
\end{equation}
where $c_{i \sigma}$ and $c_{i \sigma}^{\dag}$ are the annihilation and creation operator of electrons with spin $\sigma$ at site $i$ respectively, $t_{ij}$ is the hopping integral of electrons between site $i$ and $j$, $P_G$ is the Gutzwiller projection operator to project out doubly occupied electron states on the Cu-sites and $\left\langle ij \right\rangle$ means n.n. pairs of $i$ and $j$. Repeated spin indices are summed. As we analyzed above, the dominant orbital is $3d_{3z^2-r^2}$, thus the hopping integrals are isotropic along the x- and y- axes. And we consider only the case with n.n. and next n.n. hoppings, whose hopping integrals are $t_{nn} = t$ and $t_{nnn} = t'$ respectively.

\emph{Mean field results.} As shown in Fig.~\ref{model}(b), the superconducting order parameters of the brick-wall lattice are denoted by $\Delta_{\alpha}$ with $\alpha = 1,2,3$. Usually, the symmetry of a system will impose some constraints on the order parameters. The point group symmetry of the brick-wall lattice is $\mathcal{D}_2$ which has four irreducible representations as listed in Table \ref{char}. Since we are only interested in spin singlet pairing, only the $A$ and $B_1$ representations are relevant. For $A$ representation, We have real pairings with $\Delta_1 = \Delta_2$. For $B_1$ representation (will be denoted as $B$ hereafter), we have real pairings with $\Delta_1 = - \Delta_2, \Delta_3 = 0 $.

\begin{figure}
\centering
\includegraphics[width=\columnwidth]{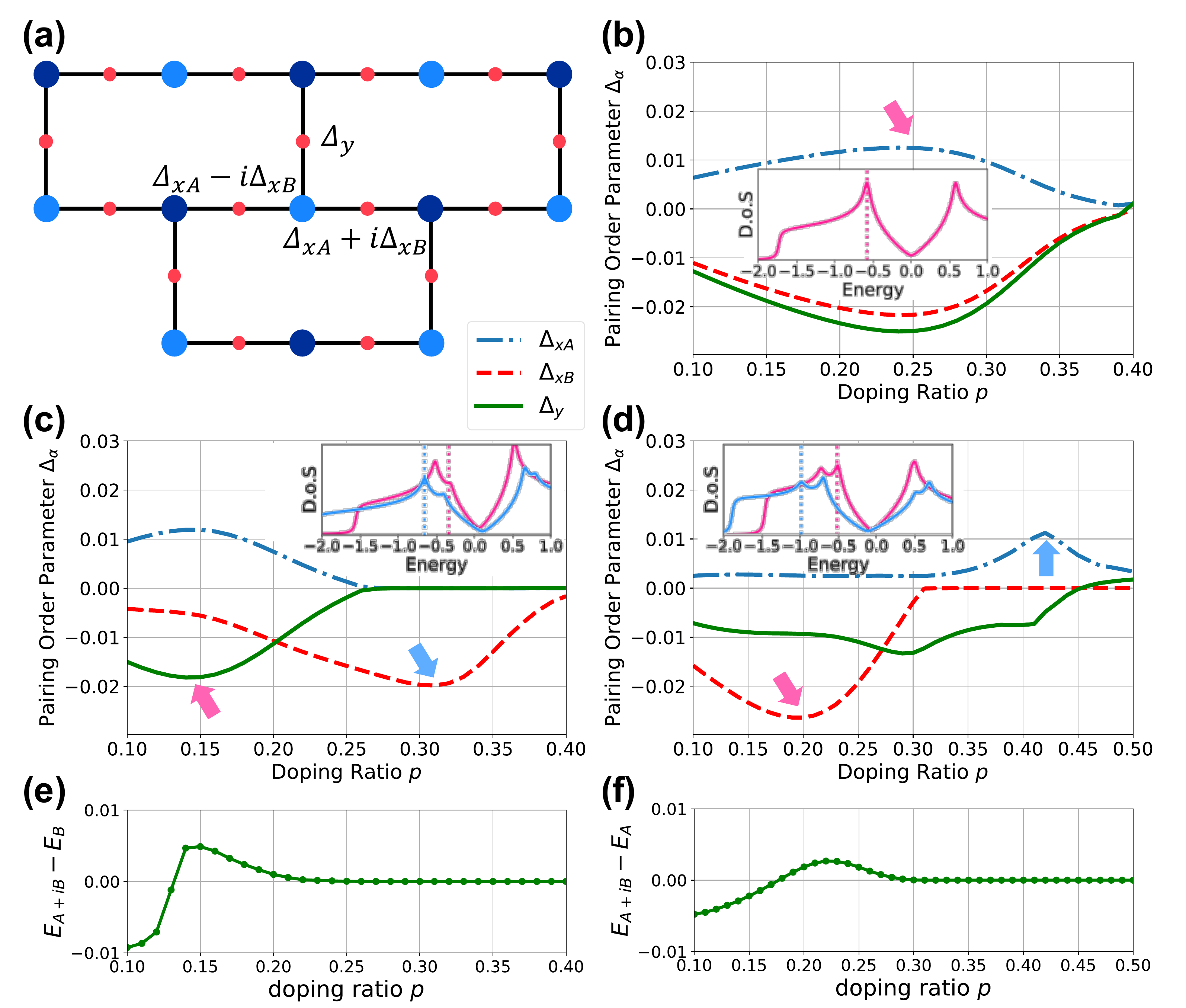}
\caption{The RMFT result of $A+iB$ phase. $\Delta_{xA}(\Delta_{xB})$ is used to indicate the contribution from $A$($B$) representation, respectively. The pairing order parameters on three distinct bonds are indicated in (a). $\Delta_{xA}$, $\Delta_{xB}$ and $\Delta_y$ obtained from the renormalized mean field theory for $t'=0$ (b), $t'/t=0.1$ (c) and $t'/t=-0.1$ (d) with $J/t=0.4$ for all cases.   The insets in (b-d) are the DOS and the corresponding Fermi energy at the optimal dopings marked by the arrows. The energy difference between the $A+iB$ phase and the pure $B$/$A$ phase for parameters of (c)/(d) are shown in (e)/(f), respectively.}
\label{result}
\end{figure}

\begin{table}
\centering
\caption{character table of group $\mathcal{D}_2$}
\begin{tabular}{c|cccc|}
Reps & e & $C_2(z)$ & $\sigma_x$ & $\sigma_y$\\
\hline
$A$ & +1 & +1 & +1 & +1\\
$B_1$ & +1 & +1 & -1 & -1\\
$B_2$ & +1 & -1 & +1 & -1\\
$B_3$ & +1 & -1 & -1 & +1\\
\hline
\end{tabular}
\label{char}
\end{table}

 $t'=0$ is a special case, where the Hamiltonian has a higher symmetry than the lattice, the $D_6$ symmetry. This can be understood as the following. The brick-wall lattice can be treated as a squeezed honeycomb lattice. If one considers only the n.n. coupling, the model can be mapped onto a t-J model with the n.n. hopping on an honeycomb lattice, and has a $D_6$ symmetry (see the details in supplementary material \footnote{See Section S1 of Supplemental Material at [URL will be inserted by publisher] for details of the mapping between the two lattices.}). Previous studies\cite{PhysRevB.81.085431,Nandkishore2012,PhysRevLett.100.146404,PhysRevB.87.094521,PhysRevB.85.035414,PhysRevB.86.020507} on t-J model on honeycomb lattice have suggested a $d+id$-wave topological superconducting phase which corresponds to a two-dimensional representation of $D_6$ group.  Although the $D_6$ symmetry is reduced to $D_2$ as one turns on the next n.n. hopping, it is possible that the pairing symmetry may still be $d+id$, or more precisely $A+iB$ in the $D_2$ case, where the 2D representation reduces to $A\oplus B$, for small $t'$. In the following calculations, we will take the $A+iB$ ansatz $\Delta_1 = \Delta_2^{*} = \Delta_{xA} + i \Delta_{xB}$ and $\Delta_3 = \Delta_y$ where $\Delta_{xA}$, $\Delta_{xB}$, and $\Delta_y$ are all real, as shown in Fig.~\ref{result}(a). $\Delta_{xA}$ and $\Delta_{xB}$ tracks the contribution from $A$ and $B$ respectively, a case with finite $\Delta_{xA}$ and vanished $\Delta_{xB}$ corresponds to pure $A$ phase, while a case with finite $\Delta_{xB}$ and $\Delta_{xA}=\Delta_y = 0$ corresponds to pure $B$ phase.

\begin{figure}
\centering
\includegraphics[width=\columnwidth]{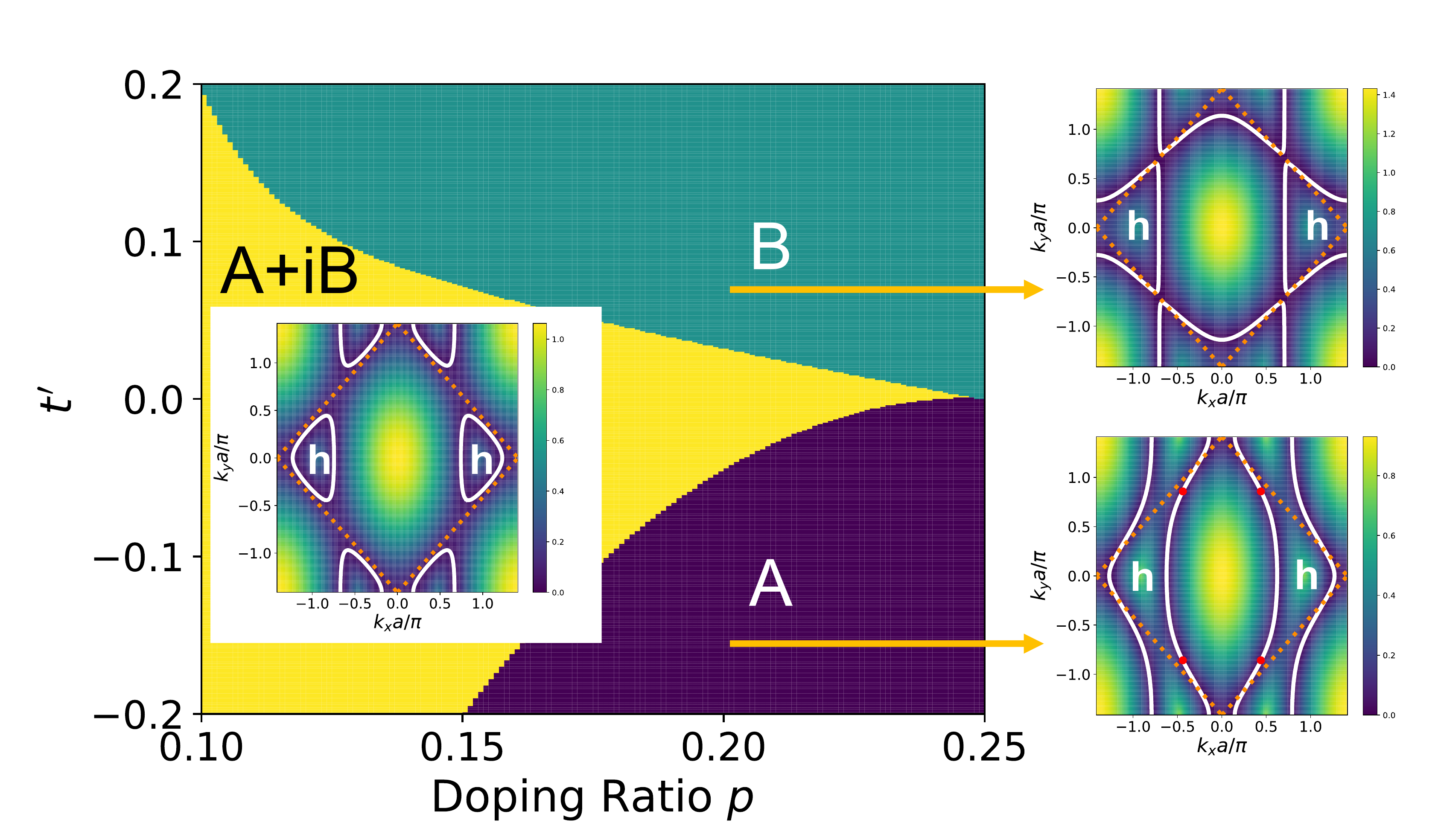}
\caption{Phase diagram of different pairing symmetry obtained from RMFT calculation.  The yellow, green, and purple Regions corresponding to $A+iB$, pure $B$, and pure $A$ phase, respectively.  We also depict the typical fermi surface topology and quasi particle gap in Brillouin zone for each phase ($t'=0$ $p=0.15$ for $A+iB$ phase, $t'=0.1t$ $p=0.31$ for $B$ phase, and $t'=-0.1$ $p=0.36$ for $A$ phase).  The orange dotted lines are the boundary of the reduced Brillouin Zone, while the symbol h indicates the region occupied by holes. The $A+iB$ and $B$ phases are gapped, while $A$ phase is gapless with its nodes depicted as red spots in the figure.}
\label{phasediagram}
\end{figure}

Then we solve the superconducting order parameters with the standard renormalized mean field theory approach (see the details in supplementary material \footnote{See Section S2 of Supplemental Material at [URL will be inserted by publisher] for details of renormalized mean field calculations.}), and the results for $t'=0$, $t'=0.1$, and $t'=-0.1$ are depicted in Fig.~\ref{result} (b), (c) and (d) respectively. The result for $t'=0$ shows a time-reversal symmetry breaking $A+iB$-wave ($d+id$-wave) phase in large doping regime, which is consistent with the previous studies on honeycomb lattice. The optimal doping of the superconducting dome corresponds to the van Hove singularities as shown in the inset of Fig.~\ref{result}(b). For small but finite $t' = \pm 0.1 t$, we still have the $A+iB$ phase at low doping ($p < 0.15$). But with increasing doping, the $A+iB$ phase becomes energetic unfavorable and is replaced by a pure $B$ phase (for $t' = 0.1t$) or a pure $A$ phase (for $t' = -0.1t$) (see fig.2(e) and (f)). In each phase, we find that the doping with maximum superconducting order parameter corresponds to the peak in density of states, i.e. the van Hove singularities.

We also perform the calculations for various dopings and $t'$, the resultant phase diagram is shown in Fig.~\ref{phasediagram}. One can find a robust $A+iB$ phase in rather large parameter regime and a phase transition from $A+iB$ phase to $A$ phase in $t'<0$ region and a transition from $A+iB$ to $B$ phase in $t'>0$ region. The different behavior between positive and negative $t'$ can be understood from the Fermi surface geometry. In Fig.~\ref{phasediagram}, we depict the typical Fermi surface for each phases. In large doping regime, the underlying Fermi surface for $t' > 0$ has a very good nesting along $k_x$ direction that favors one dimensional instability, while the Fermi surface for $t' < 0$ does not have such a nesting and favors a two dimensional instability. On the other hand, the superconducting pairing in $B$ phase is one dimensional (because $\Delta_{y}=0$), while the pairing in $A$ phase is more two dimensional. Thus one have the $B$ phase at $t'>0$ and $A$ phase at $t'<0$ in large doping regime.

As we analyzed above, the $A+iB$ phase at $t' = 0$ corresponds to the $d+id$ superconductivity, thus the $A+iB$ phase should also be a topological superconducting phase. To confirm that, we calculated the Chern number $\nu_{\text{Ch}}$ associated with each phase. We find that the $A+iB$ phase has a Chern number $\nu_{\text{Ch}}=4$, while the pure $A$ and $B$ phase is topologically trivial and has Chern number $\nu_{\text{Ch}}=0$. Thus, the phase transition between them is indeed a topological phase transition associated with a gap closing behavior.

\emph{Effects of bond disorder. }It is natural to expect the existence of bond disorder in the system that deviates the structure from a perfect brick-wall lattice, and its effects on the superconductivity need to be investigated.

\begin{figure}
\centering
\includegraphics[width=\columnwidth]{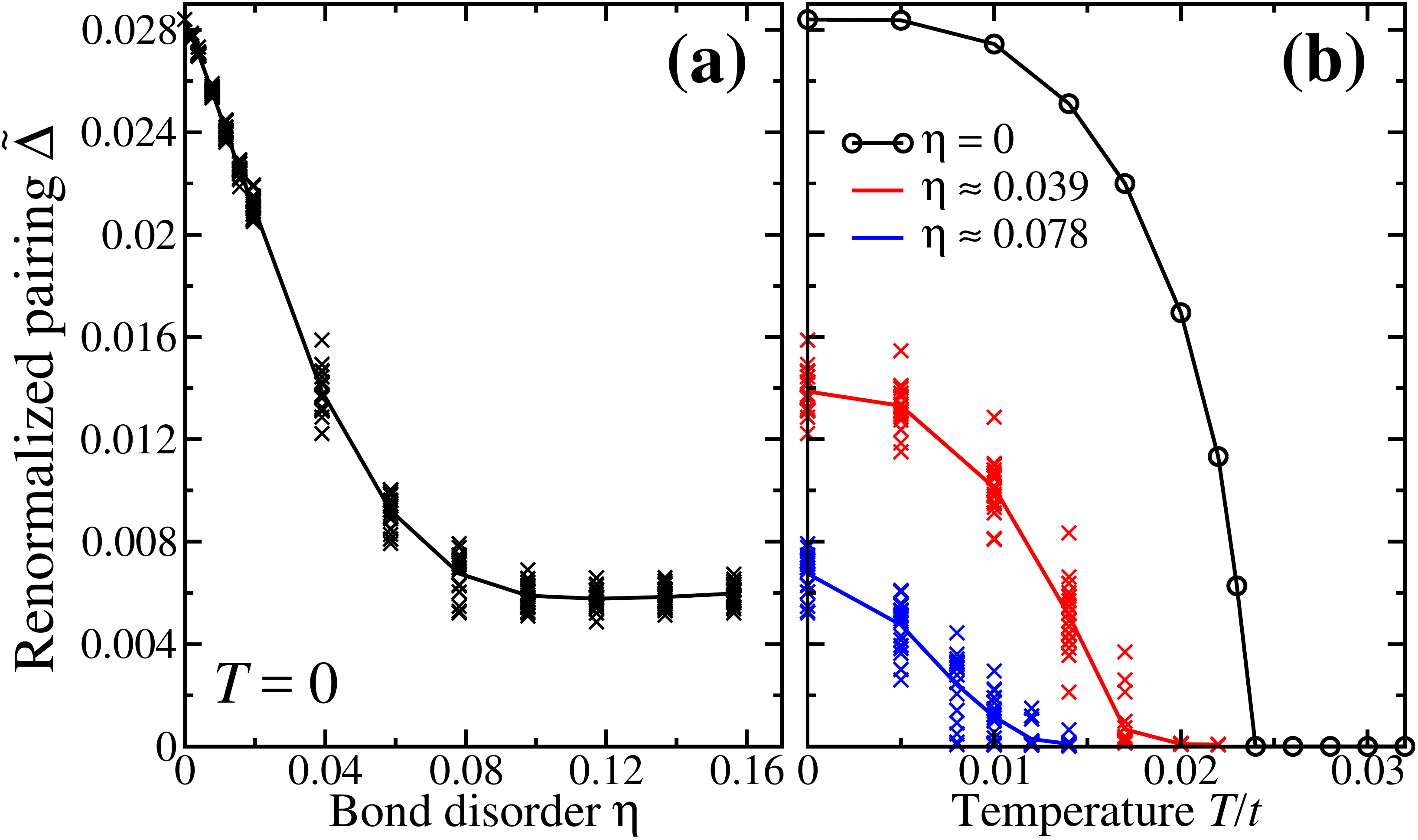}
\caption{(a) Ensemble averaged renormalized pairing $\tilde{\Delta}$ as a function of bond disorder strength $\eta$ at zero temperature. (b) Temperature dependence of the ensemble averaged $\tilde{\Delta}$ at bond disorder $\eta =0$, $\eta \simeq 0.039$, and $\eta \simeq 0.078$. Crosses represent data of each disorder realizations. }
\label{disorder}
\end{figure}

We start with a $N=32\times 32$ brick-wall lattice of periodic boundary condition, in which there are $3N/2$ n.n. bonds with nonzero hopping integral $t$ and superexchange $J$, forming the structure depicted in Fig.~\ref{model}(b). Note that, as compared with the square lattice, there are $N/2$ n.n. bonds missing in the brick-wall lattice. To introduce bond disorder, $N_\text{dis}$ n.n. bonds are redistributed randomly. Explicitly, we take out $N_\text{dis}$ bonds randomly from the $3N/2$ n.n. bonds of the brick-wall lattice and then distribute them randomly to the place of $N/2$ previously missing bonds. Clearly, $N_\text{dis}$ takes an integer value between 0 and $N/2$, and the strength of bond disorder is measured by the value of $\eta =2N_\text{dis}/N$. At a given $\eta$, we generate 20 disorder realizations and obtain the ground state of each realization self-consistently. We stay with $t'=0$ and doping concentration $\rho =0.25$ where the superconducting brick-wall lattice is in the $A+iB$ phase. The average renormalized pairing of each disorder realization, $\tilde{\Delta} = \frac{2}{3N} \sum_{\langle i,j \rangle} g^t_{ij} |\Delta_{ij}|$, at zero temperature is plotted in Fig.~\ref{disorder}(a) as crosses at various disorder strength $\eta$, with the solid line represent their ensemble averages. The ensemble averaged $\tilde{\Delta}$ decreases linearly as the disorder strength $\eta$ increases, and saturates to a nonzero minimum value at $\eta \simeq 0.09$. We thus expect the superconductivity on the perfect brick-wall lattice to survive weak bond disorders. To estimate the effects of bond disorder on the superconducting transition temperature $T_c$, the temperature dependence of $\tilde{\Delta}$ is plotted in Fig.~\ref{disorder}(b) for $\eta \simeq 0.038$ and $\eta \simeq 0.078$, with the result in the absence of disorder ($\eta =0$) is also shown for comparison. Clearly, the suppression of the mean-field critical temperature by bond disorder is much weaker than the suppression of the ensemble averaged $\tilde{\Delta}$ at zero-temperature.

\begin{figure}
\centering
\includegraphics[width=\columnwidth]{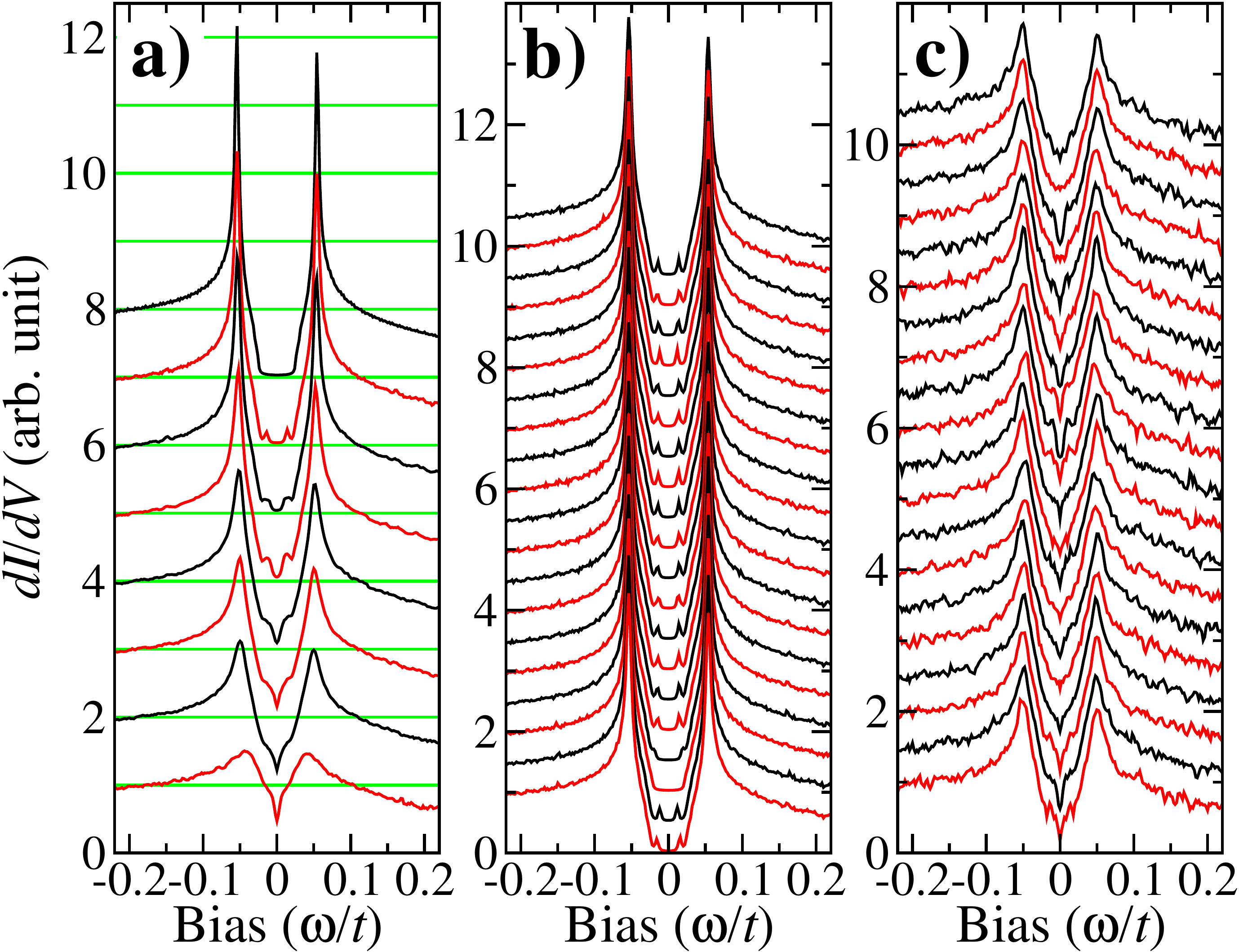}
\caption{(a) Evolution of the ensemble averaged DOS as a function of the bond-disorder strength $\eta = 2N_\text{dis}/N$ with, from top to bottom, $N_\text{dis}=$0, 1, 2, 4, 6, 8, 10, and 20. (b) The individual DOS of 20 disorder realizations with $N_\text{dis}=1$. (c) Same as (b) but for $N_{dis} = 10$. All curves are offset vertically for clarity.}
\label{disorder2}
\end{figure}

Fig. \ref{disorder2}(a) shows the evolution of the DOS, ensemble-averaged over 20 disorder realizations, as a function of the bond-disorder strength $\eta$, with the DOS of each disorder realization provided in Fig. \ref{disorder2}(b) and (c) for, respectively, $N_{dis} = 1$ and $N_{dis} = 10$. Clearly, the redistribution of a single n.n. bond in the brick-wall lattice can already produces some in-gap states. This implies that the effect of bond-disorder is more than providing a scattering potential. Furthermore, the full superconducting gap in the clean $A+iB$ phase closes at roughly $N_{dis}=8$, evident by a nonzero DOS at the Fermi energy. It would be interest to study the topology of the superconducting state with bond-disorder.

\emph{Conclusion. }In summary, we have proposed an effective brick-wall $t-t'-J$ model for the recently discovered high Tc superconductor. By using renormalized mean field theory, we have demonstrated that the superconductivity extends to very large hole concentration, and the pairing order parameter is peaked at larger hole concentration. The pairing symmetry can be complex and breaks time reversal invariance, similar to the superconductivity theoretically studied for the $t-J$ model on the honeycomb lattice. The broken of time reversal symmetry and the chiral edge states originated from its topological nature may be observed with various techniques, such as $\mu$SR, Kerr effect, SQUID etc. We note that Hubbard model on square lattice has recently been studied using more sophisticated numerical methods.  Our study based on renormalized mean field theory on the brick-wall lattice may be viewed as a starting point for these more advanced numerical methods. Our mean field result on the bond disorder effect suggests a relatively weaker reduction to the superconductivity transition temperature.

\emph{Acknowledgments.} We wish to thank C. Q. Jin and Q. Z. Huang for many useful discussions on the experimental results, Congcong Le for the density functional theory calculations to estimate the energy of the local electronic configurations. We would like also to thank J. P. Hu, Z. Q. Wang, K Jiang for many stimulating discussions. This work is in part supported by the National Natural Science Foundation of China (No. 11674151, 11847612, 11974362 and 11674278), National Key Research and Development Program of China (Grant No. 2016YFA0300300), the Strategic Priority Research Program of CAS (No. XDB28000000) and Beijing Municipal Science and Technology Commission (No. Z181100004218001).

\bibliography{BaCuO}

\end{document}